\documentclass[a4paper, times, 10pt, twocolumn, twoside,top=3.5cm,bottom=3.5cm,left=2.5cm,right=2cm]{article}

\usepackage{ISARC}

\usepackage{lscape}
\usepackage{hologo}

\begin{document}

\linespread{0.5}

\title{Constraint Control of a Boom Crane System}

\author{Michele Ambrosino$^{a}$, Arnaud Dawans$^b$, Emanuele Garone$^{a}$}

\affiliation{
$^a$Service d’Automatique et d’Analyse des Systèmes, Universitè libre de Bruxelles, Brussels, Belgium\\
$^b$Entreprises Jacques Delens S.A., Brussels, Belgium
}

\email{
\href{mailto:e.author1@aa.bb.edu}{Michele.Ambrosino@ulb.ac.be}, 
\href{mailto:e.author1@aa.bb.edu}{adawans@jacquesdelens.be},
\href{mailto:e.author1@aa.bb.edu}{egarone@ulb.ac.be}
}

\maketitle 
\thispagestyle{fancy} 
\pagestyle{fancy}

\begin{abstract}
Boom cranes are among the most used cranes to lift heavy loads. Although fairly simple mechanically, from the control viewpoint this kind of crane is a nonlinear underactuated system which presents several challenges, especially when controlled in the presence of constraints. 
To solve this problem,  we propose an approach based on the Explicit Reference Governor (ERG), which does not require any online optimization, thus making it computationally inexpensive. The proposed control scheme is able to steer the crane to a desired position ensuring the respect of limited joint ranges, maximum oscillation angle, and the avoidance of static obstacles. 
\end{abstract}

\begin{keywords}
Boom crane; Constrain control; Obstacle avoidance; Explicit reference governor; underactuated system.
\end{keywords}

\section{Introduction}\label{sec:Introduction}
Cranes are one of the most commonly used devices to hoist heavy equipment and/or materials. Due to benefits such as high maneuverability and low costs, boom cranes are particularly common. Compared with gantry cranes and tower cranes, boom cranes are much more flexible and can be easily transported and deployed. Currently, this kind of cranes are operated manually. The piece is moved in proximity of its final position, then, a worker uses his hands to finish the positioning. Considering the potentials dangers and uncertain factors of manual operations, it is essential to design efficient control methods to improve the control performance of boom cranes and restrict the payload swing amplitudes.

\medskip

Compared with gantry cranes, that are simpler and have been extensively studied in the literature, boom cranes have much more coupled and nonlinear dynamics \cite{boomcntr}-\cite{kncntr}. In particular, boom cranes involve pitching and rotational movements, which generate complicated centrifugal forces, and consequently, make the equations of motion highly nonlinear. Furthermore, as all cranes, boom cranes are underactuated \cite{und2}, having fewer independent actuators than the system degrees of freedom (DoFs). Accordingly, it is fairly challenging to control boom cranes effectively.

\medskip

In the literature, some interesting and meaningful solution have been proposed for the control of boom cranes\cite{rev}.
Open loop control schemes have been widely used because they are easy to implement. Input shaping is one of the most used open loop techniques based on a linear system. \cite{Samin}  discusses three types of input shapers  including positive and modified specified negative amplitude, positive zero vibration, and positive zero-vibration-derivative input shapers, which can reduce the sway angle during rotation. \cite{boom1} proposes a combination of input shaping and feedback control to counteract the effect of the wind. Open-loop trajectory planning methods \cite{Uchiyama}-\cite{Terashima} such as the S-curve trajectory and the straight transfer transformation model, are proposed and demonstrated to be effective for boom crane systems. The main drawback of open loop control schemes is that they are sensitive to external disturbances and to model mismatch.

\medskip

Recently some research focused on the development of closed-loop control schemes for boom cranes. Closed-loop control methods allow for increased robustness and can usually lead to better control performance in the presence of perturbations. A Linear Quadratic Regulator (LQR) is used in \cite{boom2} with a cameras system to move the crane to the desired position and reduce the payload swing angles. To reduce the payload oscillation of boom cranes, \cite{Masould} proposes a delayed position feedback antiswing control strategy. In \cite{9} the authors present a partial-state feedback control method with an integrator to achive accurate rotary positioning and swing suppression. 
A Model Predictive Control (MPC) for an industrial boom crane is shown in \cite{boom3}. A second-order sliding mode control law is proposed in \cite{boom4} for trajectory tracking and anti-sway control. In addition to model-based controllers, a series of intelligent algorithms, such as neural networks \cite{11} and fuzzy logic control \cite{12}, are also introduced for boom cranes to improve the overall control performances.

\medskip

Note that almost all existing control approaches focus on stabilization objectives, i.e., boom positioning and payload swing suppression. However, from the practical perspective, the payload swing’s transient responses also need to be ensured, e.g., swing angles should be restricted within prescribed safety ranges to ensure stable transportation. Furthermore, it must also be guaranteed that the trajectories avoid obstacles (e.g. walls).

\medskip

On the basis of this analysis it is possible to point out two open problems in the control of boom cranes: 

\begin{itemize}[noitemsep]
    \item Existing closed-loop control laws focus on reducing residual oscillations only when the system reaches the desired position. Instead, a good control strategy should take into account of oscillations throughout the movement in order to reduce the energy stored by the system and avoid potentially dangerous situations;
    \item There are only few results for the trajectory planning (and for obstacle avoidance) for this kind of crane (see e.g. \cite{18}). These solutions are almost always very complicated, based on the specific nonlinear trajectory planning process and focus on kinematics aspects, not taking into account the dynamics of the crane.
\end{itemize}


Recently a novel approach for the control of nonlinear systems subject to constraints called Explicit Reference Governor (ERG) has been introduced in \cite{13}-\cite{19}. The ERG is an add-on control unit to be used on a pre-stabilized system and is based on the general Reference Governor (RG) philosophy (see \cite{GARONE14} for a survey on RG schemes) which ensures constraint satisfaction by manipulating the reference of a pre-stabilized system so that the transient response does not violate the constraints. An interesting feature of the ERG is that it can enforce both state and input constraints of nonlinear systems without having to solve an online optimization problem.

\medskip

This paper \textit{aims} at designing a novel control framework, the ERG, for boom crane subject to limited joint ranges and static obstacle avoidance constraints.

\section{Dynamic Model and Problem Statement}

The dynamic model of a 3-D boom crane (see Fig.~\ref{fig:boomcrane}), with fixed cable length, can be described by the following equations \cite{Uchiyama}:
\begin{figure}[ht!]
\centering
\includegraphics[width=0.7\columnwidth]{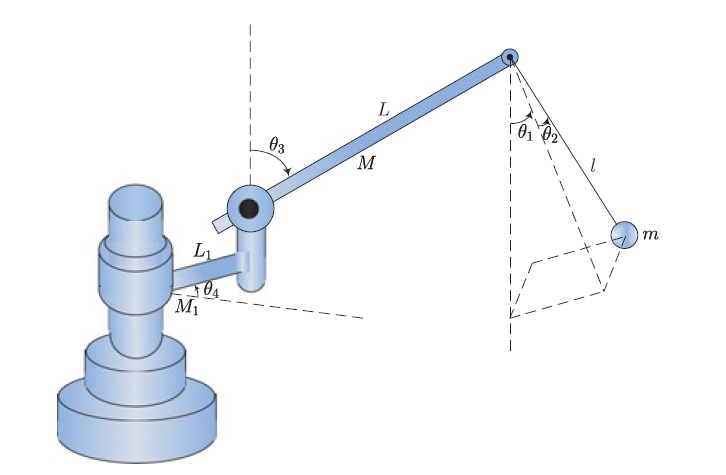}
\caption{\label{fig:boomcrane} Model of an underactuated boom crane system. \cite{20}}
\end{figure}

\begin{equation} \label{eq:model1}
\begin{array}{ll}
ml^2(1+\theta_1^2)\ddot\theta_1+ml^2\theta_1\theta_2\ddot\theta_2+mlL(-\theta_1sin\theta_3+cos\theta_3)\ddot\theta_3 \\[1ex] 
-ml^2\theta_2\ddot\theta_4+ml^2\theta_1(\dot\theta_1^2\dot\theta_2^2) -mlL(sin\theta_3+\theta_1cos\theta_3)\dot\theta_3^2 \\[1ex] 

-ml(l\theta_1+Lsin\theta_3)\dot\theta_4^2-2ml^2\dot\theta_2\dot\theta_4+mgl\theta_1 = 0,
\end{array}
\end{equation}

\begin{equation} \label{eq:model2}
\begin{array}{ll}
ml^2\theta_1\theta_2\ddot\theta_1+ml^2(1+\theta_2^2)\ddot\theta_2-mlL\theta_2sin\theta_3\ddot\theta_3 \\[1ex] 
+(ml^2\theta_1+mlLsin\theta_3)\ddot\theta_4+ml^2\theta_2(\dot\theta_1^2+\dot\theta_2^2) \\[1ex] 
-mlL\theta_2cos\theta_3\dot\theta_3^2-ml^2\theta_2\dot\theta_4^2+2ml^2\dot\theta_1\dot\theta_4+2ml^2\dot\theta_1\dot\theta_4 \\[1ex] 
+2mlL\dot\theta_3\dot\theta_4cos\theta_3+mgl\theta_2 = 0,
\end{array}
\end{equation}

\begin{equation} \label{eq:model3}
\begin{array}{ll}
mlL(cos\theta_3-\theta_1sin\theta_3)\ddot\theta_1-mlL\theta_2sin\theta_3\ddot\theta_2 \\[1ex] 
+(mL^2+J_y)\ddot\theta_3-mlL\theta_2cos\theta_3\ddot\theta_4-mlLsin\theta_3(\dot\theta_1^2+\dot\theta_2^2) \\[1ex] 
-[\frac{1}{2}(J_x-Jz)sin(2\theta_3)+mlL\theta_1cos\theta_3+\frac{1}{2}mL^2sin(2\theta_3)]\dot\theta_4^2 \\[1ex] 
-2mlL\dot\theta_2\dot\theta_4cos\theta_3-g(\frac{1}{2}ML+mL-\frac{1}{2}M_1L_1)sin\theta_3 = u_3,
\end{array}
\end{equation}

\begin{equation} \label{eq:model4}
\begin{array}{ll}
-ml^2\theta_2\ddot\theta_1+(ml^2\theta_1+mlLsin\theta_3)\ddot\theta_2-mlL\theta_2cos\theta_3\ddot\theta_3 \\[1ex] 
+[mL^2(sin\theta_3)^2+ml^2(\theta_1^2+\theta_2^2)+2mL\theta_1sin\theta_3+I_b \\[1ex] 
+J_x(sin\theta_3)^2+Jz(cos\theta_3)^2]\ddot\theta_4+[mL^2dot\theta_3sin(2\theta_3) \\[1ex] 
+2ml^2(\theta_1\dot\theta_1+\theta_2\dot\theta_2)+2mL(\dot\theta_1sin\theta_3+\theta_1\dot\theta_3cos\theta_3) \\[1ex] 
+(J_x-J_z)\dot\theta_3sin(2\theta_3)]\dot\theta_4 + mlL\theta_2\dot\theta_3^2sin\theta_3 = u_4.
\end{array}
\end{equation}

The system parameters are reported in Tab.~\ref{tb:par}. The system can be compactly rewrite as

\begin{equation}\label{eq:modelmatrix}
{{M(q)\ddot{q} + C(q,\dot{q})\dot{q} + g(q)} = {\begin{bmatrix}
0_{2x2} \\ I_{2x2} \end{bmatrix}}u, }
\end{equation}

where q = $[\theta_1, \theta_2, \theta_3, \theta_4]^T \in{\mathbb{R}^4}$ represents the state vector, and u = $[u_3, u_4]^T \in{\mathbb{R}^2}$ represents the control input vector. The matrices M(q) $\in{\mathbb{R}^{4x4}}$, $ C(q,\dot{q})\in{\mathbb{R}^{4x4}}$, and g(q) $\in{\mathbb{R}^{4}}$ represent the inertia, centripetal-Coriolis, and gravity.

\begin{table}[hb]
\begin{center}
\caption{Parameters of the boom crane system}\label{tb:par}
\begin{tabular}{cccc}
Parameters & Physical & Units \\\hline
$\theta_1(t)$ & Payload radial swing angle & rad\\
$\theta_2(t)$ & Payload tangential swing angle & rad \\ 
$\theta_3(t)$ & Boom pitch angle & rad \\ 
$\theta_4(t)$ & Boom yaw angle & rad \\ 
M & Boom mass & kg \\ 
m & Payload mass & kg \\ 
$M_1$ & Ballast mass & kg \\ 
L & Boom length & m \\ 
l & Rope length & m \\ 
$L_1$ & Ballast length & m \\ 
$J_x,J_y,J_z$ & Moments of inertia of the boom & $kg\cdot m^2$ \\ 
$I_b$ & Ballast inertia moment & $kg\cdot m^2$ \\ 
$u_3(t),u_4(t)$ & Control inputs & $N\cdot m$ \\ \hline
\end{tabular}
\end{center}
\end{table}

\subsection{Control objective}

The control objective is to move the boom to a desired position and dampen the load swing at the same time. This can be described mathematically as follows:
\begin{equation}\label{eq:constraints}
\begin{array}{ll}
\lim_{t\to \infty}\theta_1(t) = 0, \hspace{0.3cm} \lim_{t\to \infty}\theta_2(t) = 0,  \\[2ex] 
\lim_{t\to \infty}\theta_3(t) = \theta_{3d}, \hspace{0.2cm} \lim_{t\to \infty}\theta_4(t) = \theta_{4d},
\end{array}
\end{equation}
where $\theta_{3d}$ and $\theta_{4d}$ are the boom’s desired angles. 

Typically in this kind of applications we have three main types of constraints: constraints concerning the maximum range of the joints, safety constraints related to the suspended load, and constraints modelling the collision avoidance with objects and structures (e.g. walls).  

In this paper for the joint range constraints we will assume that the  boom pitch angle (i.e., $\theta_{3}(t)$) is constrained within the range $(\frac{8\pi}{9},\frac{-\pi}{18})$. Thus

\begin{equation}\label{eq:constt31}
{\theta_3\leq{\frac{8\pi}{9}},}
\end{equation}
\begin{equation}\label{eq:constt32}
{-\theta_3\leq{-\frac{\pi}{18}}. }
\end{equation}

For what concerns safety constraints linked to the swinging load, a number of different constraints can be defined. A simple form of safety constraints is to impose that  $\theta_i$ never violates a maximum swing angle. In this paper we will consider the constraint $\left|{\theta_i}\right|\leq\theta_{mi}$, with i =1,2, where $\theta_{mi}=\frac{\pi}{36}$. Thus

\begin{equation}\label{eq:constt11}
{\theta_1\leq{\pi\over36},}
\end{equation}
\begin{equation}\label{eq:constt2}
{-\theta_1\leq{\pi\over36}, }
\end{equation}
\begin{equation}\label{eq:constt21}
{\theta_2\leq{\pi\over36}, }
\end{equation}
\begin{equation}\label{eq:constt22}
{-\theta_2\leq{\pi\over36}, }
\end{equation}

\begin{figure}[ht!]
\centering
\includegraphics[width=0.8\columnwidth]{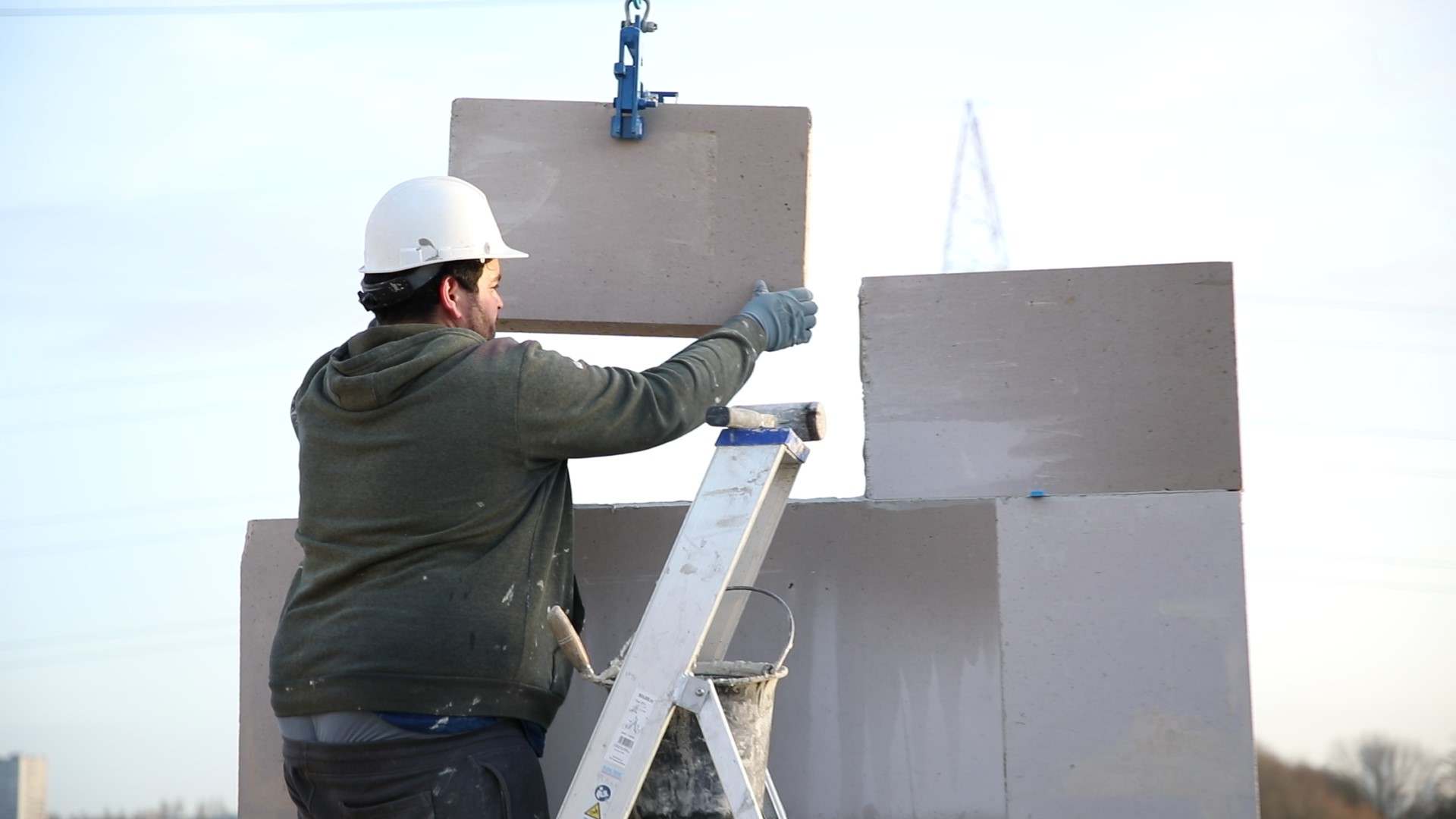}
\caption{\label{fig:mason} Mason activity}
\end{figure}

Collision avoidance constraints are  constraints where we want to avoid that the load collides with some object/structure. We consider a wall construction scenario where the mason waits for a new brick to bring it to its final position (Fig.~\ref{fig:mason}). In this paper we assume that we want to avoid the collision of the swinging load with three obstacles that are present during this activity: the mason, the brick already placed and the wall. 

\medskip

\begin{figure}[ht!]
\centering
\includegraphics[width=1\columnwidth]{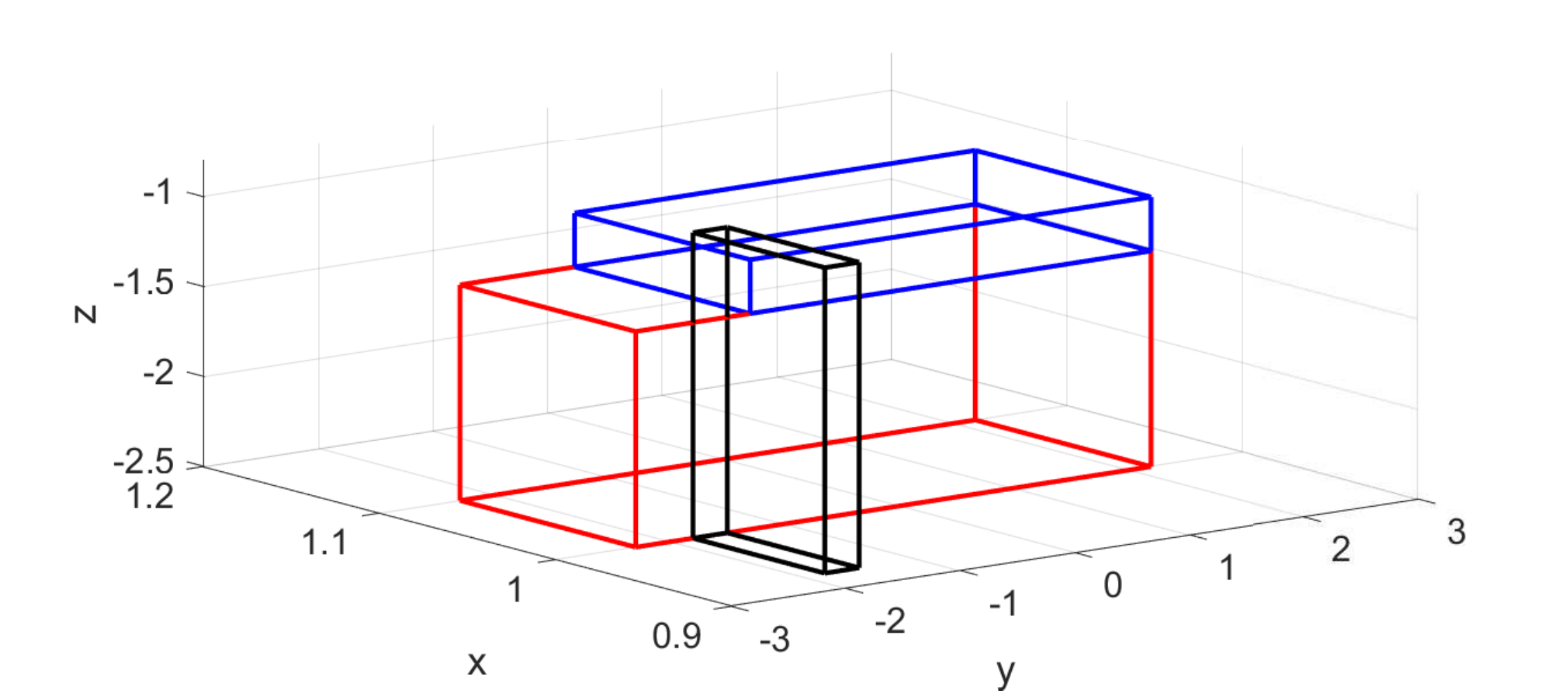}
\caption{\label{fig:wall}Obstacle constraints. Blue line: Bricks. Red line: Wall. Black line: Mason .}
\end{figure}

Since the cable length is fixed and since constraints (\ref{eq:constt11})-(\ref{eq:constt22}) will be enforced, we embed the obstacles into boxes taking into account a safety margin equal to the maximum displacement between the rest condition of the load and the maximum swing compatible with constraints (\ref{eq:constt11})-(\ref{eq:constt22}).
\newline
A schematic view of these three boxes is shown in the Fig.~\ref{fig:wall}.

To write these three constraints in the operational space, the direct kinematics of the crane is used \cite{18}:  
\begin{equation}\label{eq:x}
\begin{array}{ll}
x = Lsin(\theta_3)cos(\theta_4)+l\theta_1cos(\theta_4)-l\theta_2sin(\theta_4),
\end{array}
\end{equation}
\vspace{0.05cm}
\begin{equation}\label{eq:y}
\begin{array}{ll}
y = Lsin(\theta_3)sin(\theta_4)+l\theta_1sin(\theta_4)+l\theta_2cos(\theta_4),
\end{array}
\end{equation}
\vspace{0.05cm}
\begin{equation}\label{eq:z}
\begin{array}{ll}
z = Lcos(\theta_3)-l\theta_2cos(\sqrt{\theta_1^2+\theta_2^2}).
\end{array}
\end{equation}

\begin{figure}[ht!]
\centering
\includegraphics[width=1\columnwidth]{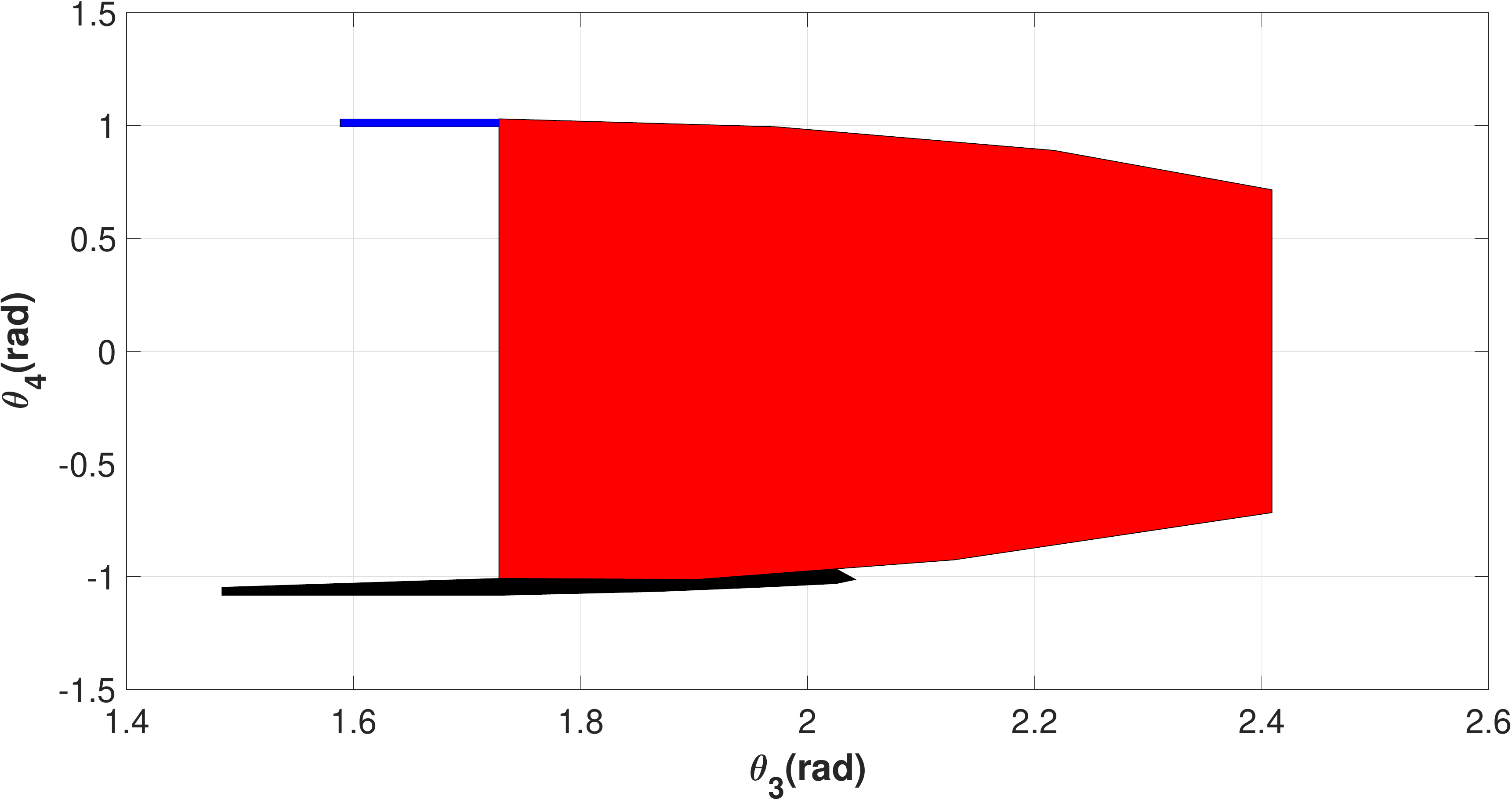}
\caption{\label{fig:embedding} The constraints in the joint space. Blue: Bricks. Red: Wall. Black: Mason. }
\end{figure}

Thanks to (\ref{eq:x})-(\ref{eq:z}) we can translate the constrains of each obstacles (e.g. mason, brick and wall) into the joint space. Since the boxes considering the constraints are robust for any swing satisfying (\ref{eq:constt11})-(\ref{eq:constt22}), we can consider $\theta_1 = \theta_2 = 0$, and write the constraints only in terms of $\theta_3$ and $\theta_4$ as follows

\begin{equation}\label{eq:wall_const}
{W_{i} = \{h_{i}(\theta_3,\theta_4)\geq0\},\quad i=1...3,}
\end{equation}

where $h_i$ is a nonlinear function. These three nonlinear constraints mapped in the joint space are reported in Fig.~\ref{fig:embedding}.
Note that in the joint space each of these constraints is easily embeddable as a union of linear constraints. This fact can be exploited for the control law following the same lines proposed in \cite{17}.

\medskip

In conclusion, constraints (\ref{eq:constt31})-(\ref{eq:constt22}) are linear constraints, while the constraints obtained by (\ref{eq:wall_const}) are non linear constraints. 

\medskip

The main goal of this paper is to build a control law able to stabilize the system around each desired point of equilibrium ensuring good dynamic performances while, at the same time, ensuring the satisfaction of constraints  (\ref{eq:constt31})-(\ref{eq:constt22}), and (\ref{eq:wall_const}).


\section{Control Desing}\label{control}

\begin{figure}[ht!]
\centering
\includegraphics[width=0.9\columnwidth]{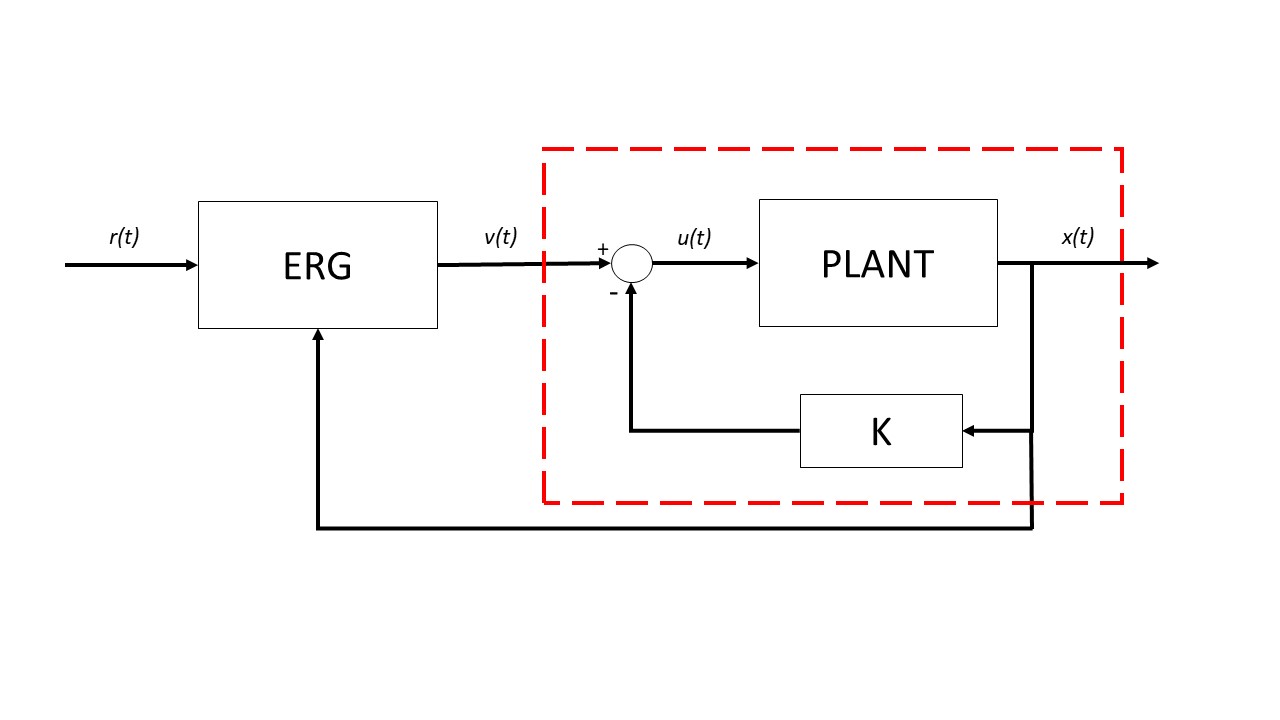}
\caption{\label{fig:block}ERG based feedback control scheme.}
\end{figure}

The control architecture proposed in this paper consists of two cascade control loops as shown in Fig.~\ref{fig:block}. The first loop pre-stabilizes the system, whereas the second loop manipulates the reference of the pre-stabilized system to ensure constraint satisfaction and reference tracking. In this paper a Linear Quadratic Regulator is used in the inner loop to ensure stability and fast dynamics. For what concerns the constraints management, the external loop consists of an Explicit Reference Governor.

\subsection{LQR Synthesis}

Let us define $x = [q\quad\dot{q}]^{T}$ as state of our system.
\newline
In order to design a control law for the system at hand we proceeded with the linearization of the nonlinear equations (\ref{eq:modelmatrix}) around the point of equilibrium $x_v = [0,0,\frac{\pi}{3},0,0,0,0,0]^T$, and considering as equilibrium input $u_{eq}$:
\begin{equation}\label{eq:ueq}
{u_{eq} = \begin{bmatrix} -g(\frac{1}{2}ML+mL-\frac{1}{2}M_1L_1)sin(x_{v,3}),  \\ 0 \end{bmatrix}}
\end{equation}
which represents the so-called "desired gravity compensation".
\newline
The resulting linearized system
\begin{equation}\label{eq:lin_mod}
{\delta \dot{x(t)} = A\delta x(t) + B\delta u(t),}
\end{equation}

is then used to compute an LQR control law $\delta u(t) = -K\delta x$ for the linear system (\ref{eq:lin_mod}). \newline
\newline
Using this gain matrix $K$ the following control law is obtained 
\begin{equation}\label{eq:u_nl}
{u = -K(x-x_v)+u_{eq},}
\end{equation}

 It is worth noticing that the matrix $A$ in (\ref{eq:lin_mod}) depends on the value of the equilibrium angle $\theta_3$. However, it has been numerically verified that, for the choices of weight used in this paper and in the operative ranges prescribed by the system constraints, the control law (\ref{eq:u_nl}) is able to stabilize the system regardless of the initial condition.

\subsection{ERG Synthesis}\label{erg}
The idea behind the Explicit Reference Governor \cite{Garone15} is to generate the applied reference signal \textit{v} (see Fig.~\ref{fig:block}) so that, if \textit{v} was to be frozen at any time instant, the transient dynamics of the pre-stabilized system would not violate the constraints. This is achieved by manipulating the derivative of the applied reference  in continuous time using the nonlinear control law.

\begin{equation}\label{eq:vdot}
{\dot v(t) = \Delta(x(t),v(t))\rho(r(t),v(t)),}
\end{equation}

with 

\begin{equation}\label{eq:delta}
{\Delta(x(t),v(t)) = k\min_i(\Gamma_i(v(t))-V_i(q(t),v(t)))},
\end{equation}

where $k>0$ in a tuning parameter, and i=1...$n_c$, with $n_c$ is the number of the constraits. $\Delta(x(t),v(t))$ and $\rho(r(t),v(t))$ are the two fundamental components of the ERG scheme, called the Dynamic Safety Margin (DSM) and the Navigation Field (NF), respectively. The scheme is proven to
ensure recursive feasibility and asymptotic convergence to a constant reference \cite{Garone15}. \newline
The definition of the DSM differs for the linear constraints and non linear constraints. 
\newline
For the linear constraint, the first step is to write the (\ref{eq:constt31})-~(\ref{eq:constt22}) in the form $\beta_{x,i}x\leq d_{i}, i = 1...6$. 
For the $i^{th}$ linear constraints (\ref{eq:constt31})-(\ref{eq:constt22}), $\Gamma(v(t))$ is 

\begin{equation}\label{eq:gammai}
{\Gamma_i(v) = {{(\beta_{x,i}^Tx_v+\beta_{v,i}^Tv-d_i)^2}\over{\beta_{x,i}^TP_i^{-1}\beta_{x,i}}},}
\end{equation}

where accordingly to \cite{13}, the matrix $Pi>0$ can be found by solving the offline LMI optimization problem:

\begin{equation}\label{eq:Popt}
\begin{array}{ll}

P_i = \min_P logdet(P) \\
s.t. \\
(A-BK)^TP+P^T(A-BK)<0 \\
P>{\beta_{x,i}^T\beta_{x,i}\over{||\beta_{x,i}||^2}}

\end{array}
\end{equation}

\begin{figure}[ht!]
\centering
\includegraphics[width=1\columnwidth]{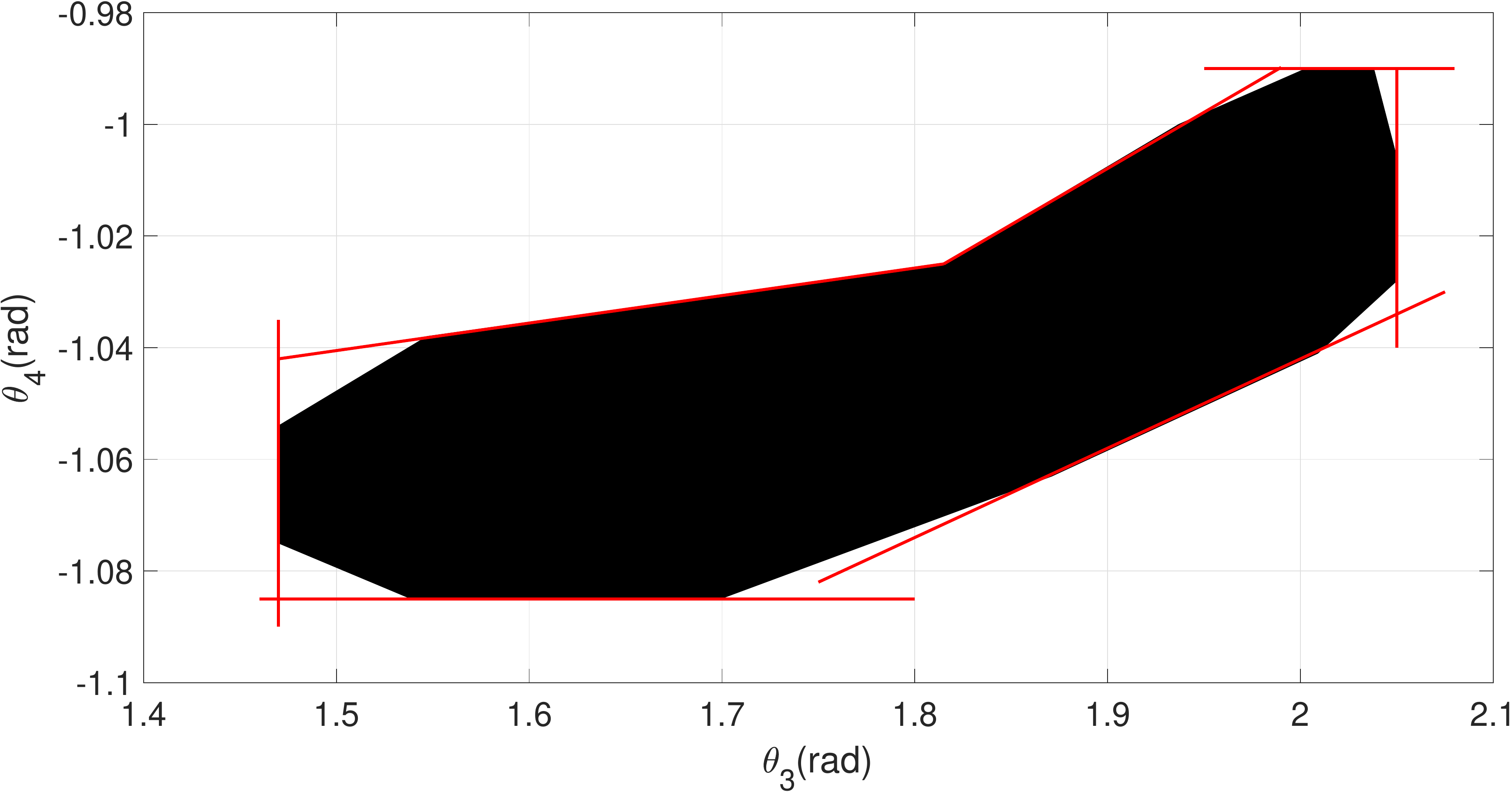}
\caption{\label{fig:tang}Embedding mason constraint}
\end{figure}

For the nonlinear constraints (\ref{eq:wall_const}), to be able to evaluate the DSM, we propose the following procedure.
\newline
These constraints are well embedded  as a union of linear constraints. Fig.\ref{fig:tang} shows the embedding for the mason constraint. The same approach is used for the other two constraints. In this way, for each obstacle, we obtain a set of  linear constraints that we can exploit in the design of the control law. Accordingly, we can rewrite each new set as the union of sets described by linear constraints in the form $\beta_{t,i,j}x\leq d_{i,j}, i = 1...n_{t,j},\quad j = 1...3$, where $n_{t,j}$ is the number of tangent used for each of the embeddings.

\medskip

Accordingly to \cite{17} the DSM of the union of $n_{t,j}$ sets defined by linear constraints can be evaluated as

\begin{equation}\label{eq:delta_conc}
{\Delta_{t,j}(q(t),v(t)) = \max_{l=1,...,n_t}(\Delta_{t,l,j}(q(t),v(t))},
\end{equation}

where

\begin{equation}
 \Delta_{t,l,j}(q,v)=\Gamma_{t,l,j}(v)-V_{t,l,j}(q,v)  ,
\end{equation}

whit l = 1,...,${n_{t,j}},\quad j = 1...3$, where $\Gamma_{t,l,j}$ and $V_{t,l,j}$ can be evaluated by solving the (\ref{eq:gammai})-(\ref{eq:Popt}) for each tangent constraint.

For what regards the NF, it can be designed by decoupling into an attraction and a repulsion term as \cite{16}

\begin{equation}\label{eq:NF}
{\rho(r(t),v(t)) = \rho_a(r(t),v(t))+\rho_r(r(t),v(t))},
\end{equation}

 where $\rho_a(r(t),v(t))$ is a vector field which points towards the desired position, and $\rho_r(r(t),v(t))$is a vector field which points away from the constraints. For the attraction term $\rho_a(r(t),v(t))$, the most intuitive choice is

\begin{equation}\label{eq:NFa}
\rho_a(r(t),v(t)) = {r(t)-v(t)\over{\max\{||r(t)-v(t)||,\eta\}}},
\end{equation}

where $\eta>0$ is a smoothing factor. Repulsion terms can be split into two terms as:

\begin{equation}\label{eq:NFr}
    \rho_r(r(t),v(t)) = \rho_{r,1}(r(t),v(t)) + \rho_{r,2}(r(t),v(t)),
\end{equation}

where

\begin{equation}
\rho_{r,1}(r(t),v(t)) =  {\displaystyle\sum_{i=1}^{6}} \max\biggl\{{{\zeta-\beta_{x,i}x(t)-d_i\over\zeta-\delta},0}\biggr\}{\beta_{x,i}\over{||\beta_{x,i}||}},
\end{equation}
\begin{equation}
    \rho_{r,2}(r(t),v(t)) = {\displaystyle\sum_{j=1}^{3}} \max\biggl\{{{\zeta-\beta_{l(t)_j^\star}x(t)-d_{l(t)_j^\star}\over\zeta-\delta},0}\biggr\}{\beta_{l(t)_j^\star}\over{||\beta_{l(t)_j^\star}||}},
\end{equation}

where  $\zeta>\delta>0$ and ${(t)^\star}$ is the index such that $\Delta_{t,j}(q(t),v(t)) = \Delta_{t,l(t)^\star,j}(q(t),v(t))$ in \ref{eq:delta_conc}.

\medskip

In this paper, we will use a discrete-time implementation of the ERG. Note that, one could use the Euler approximation of (\ref{eq:vdot}), such as

\begin{equation}\label{eq:v_discr}
{v(k+1) = v(k)+kT_s\Delta(q,v)\rho(r(k),v(k)),}
\end{equation}

where $T_s$ is the sampling time of the system. However, as it is, this approximation might not ensure recursive feasibility if $T_s$ is not sufficiently small with respect to the dynamics of the system. For this reason in this paper we will use the scheme introduced in (\cite{19}), which verifies that the candidate reference $\hat{v}(k)$ ensures recursive feasibility one step ahead. To do so, it predicts the evolution of the close loop states $\hat{x}_{cl}(k+1)$, given this candidate reference $\hat{v}(k)$, and evaluates if the dynamic safety
margin is positive when maintaining the candidate reference one step ahead, i.e., $\Delta(\hat{x}_{cl}(k+1),\hat{v}(k))$. If the candidate reference holds feasibility, it is applied as the reference at the current step, i.e. v(k) = $\hat{v}(k)$.

\section{Simulation Results}

To demonstrate the effectiveness of the proposed ERG strategy, in this section we simulate the boom crane shown in Fig.~\ref{fig:boomcrane}.The
physical parameters are shown in Tab.~\ref{tb:valpar}
\begin{table}[hb]
\begin{center}
\caption{Physical Parameters}\label{tb:valpar}
\begin{tabular}{cccc}
Parameters & Value & Units \\\hline
M & 2.5 & kg \\ 
m & 3.5 & kg \\ 
$M_1$ & 6 & kg \\ 
L & 2  & m \\ 
l & 1 & m \\ 
$L_1$ & 0.5 & m \\\hline
\end{tabular}
\end{center}
\end{table}

The parameters of the control architecture of Fig.~\ref{fig:block}, are as follows.\newline
For the inner feedback loop:
\begin{equation}
\small
{K = \begin{bmatrix}
-106.1665,0,89.6362,0,-11.28,0,68.877,0 \\
0,-91.3357,0,31.6228,0,-7.8488,0,52.3688
\end{bmatrix} }
\end{equation}
For the ERG loop, $k = 30$, $\eta = 10e^{-4}$, $\omega=0.6$,$\zeta = 10$, $\delta = 0.09$.\newline

\begin{figure}[ht!]
\centering
\includegraphics[width=1\columnwidth]{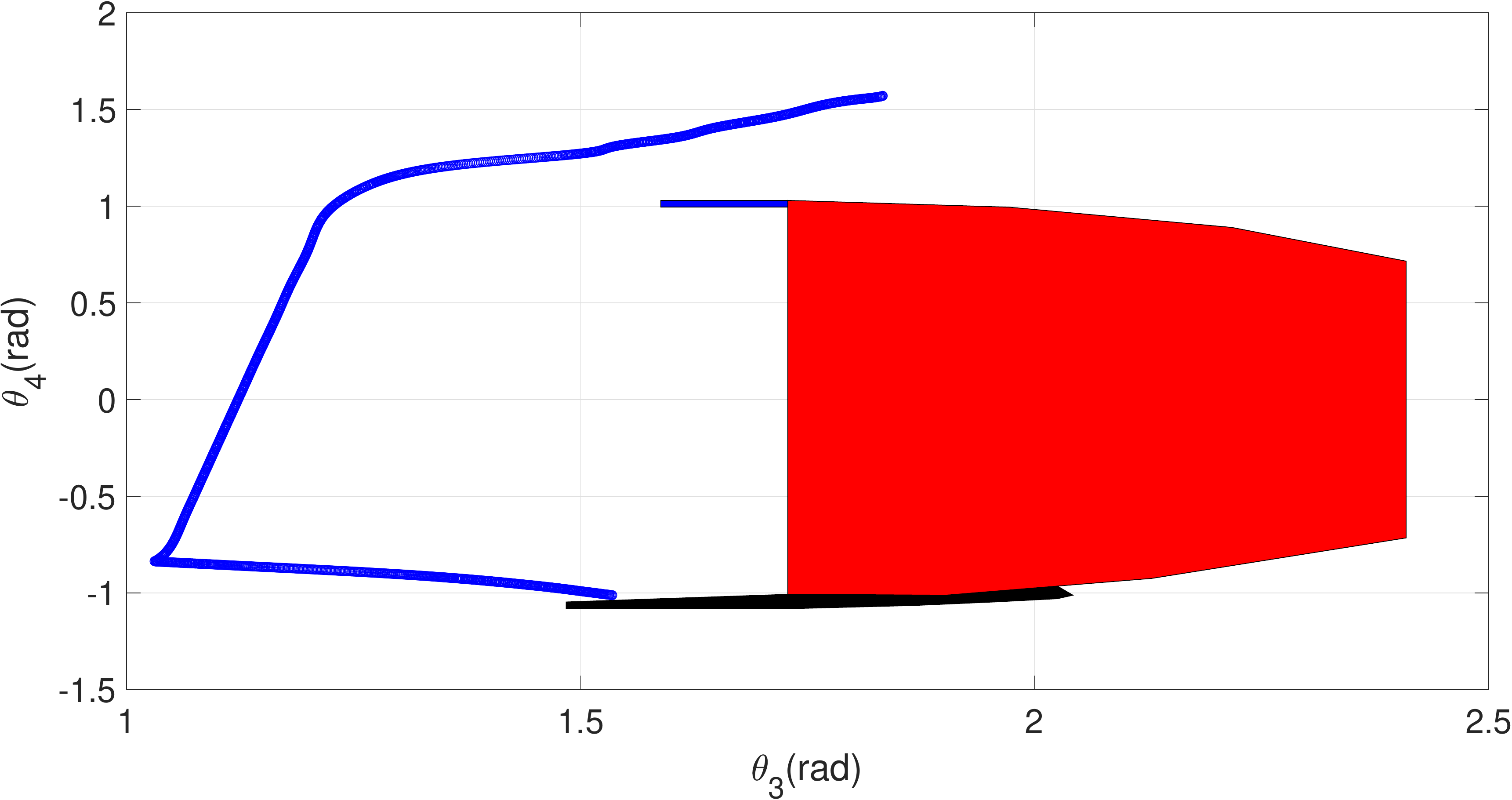}
\caption{\label{fig:ERG1} Trajectory in the joint space}
\end{figure}

\begin{figure}[ht!]
\centering
\includegraphics[width=1\columnwidth]{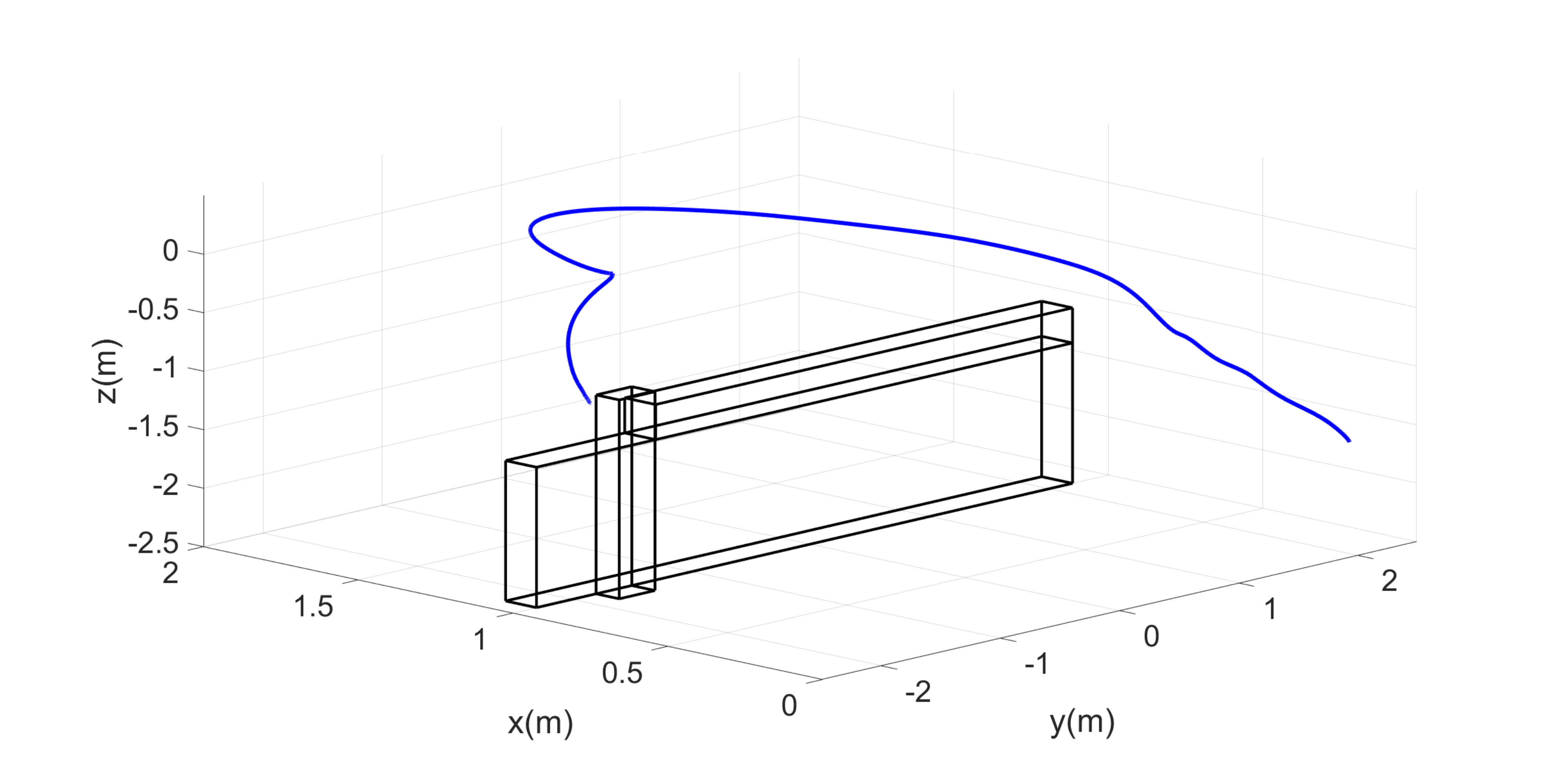}
\caption{\label{fig:ERG2} Trajectory in the operational space}
\end{figure}

\begin{figure}[ht!]
\centering
\includegraphics[width=1\columnwidth]{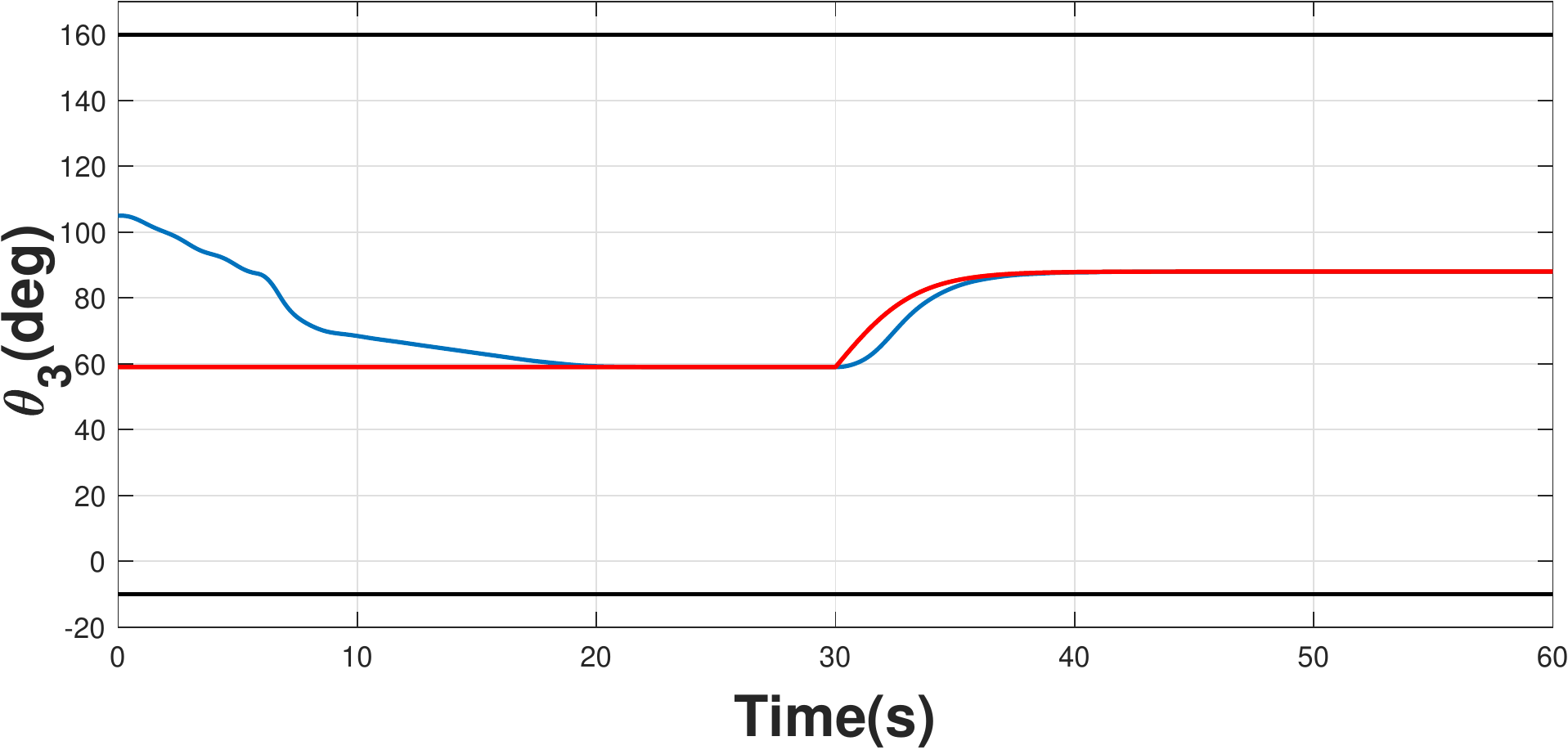}
\caption{\label{fig:t3} Boom pitch angle $\theta_3$. Red line: Desired reference. Blue line: Real value. Black line: Constraints }
\end{figure}

\begin{figure}[ht!]
\centering
\includegraphics[width=1\columnwidth]{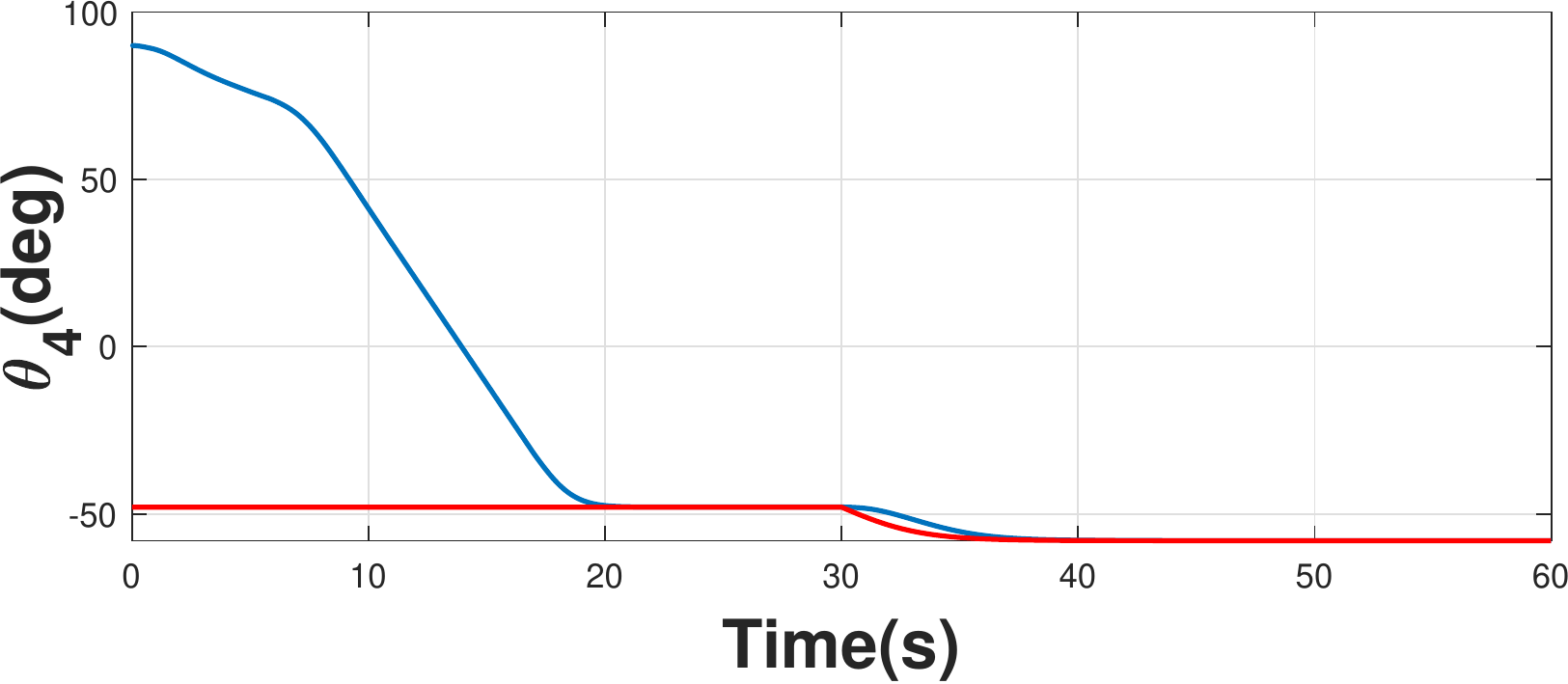}
\caption{\label{fig:t4} Yaw pitch angle $\theta_4$. Red line: Desired reference. Blue line: Real value. }
\end{figure}

\begin{figure}[ht!]
\centering
\includegraphics[width=1\columnwidth]{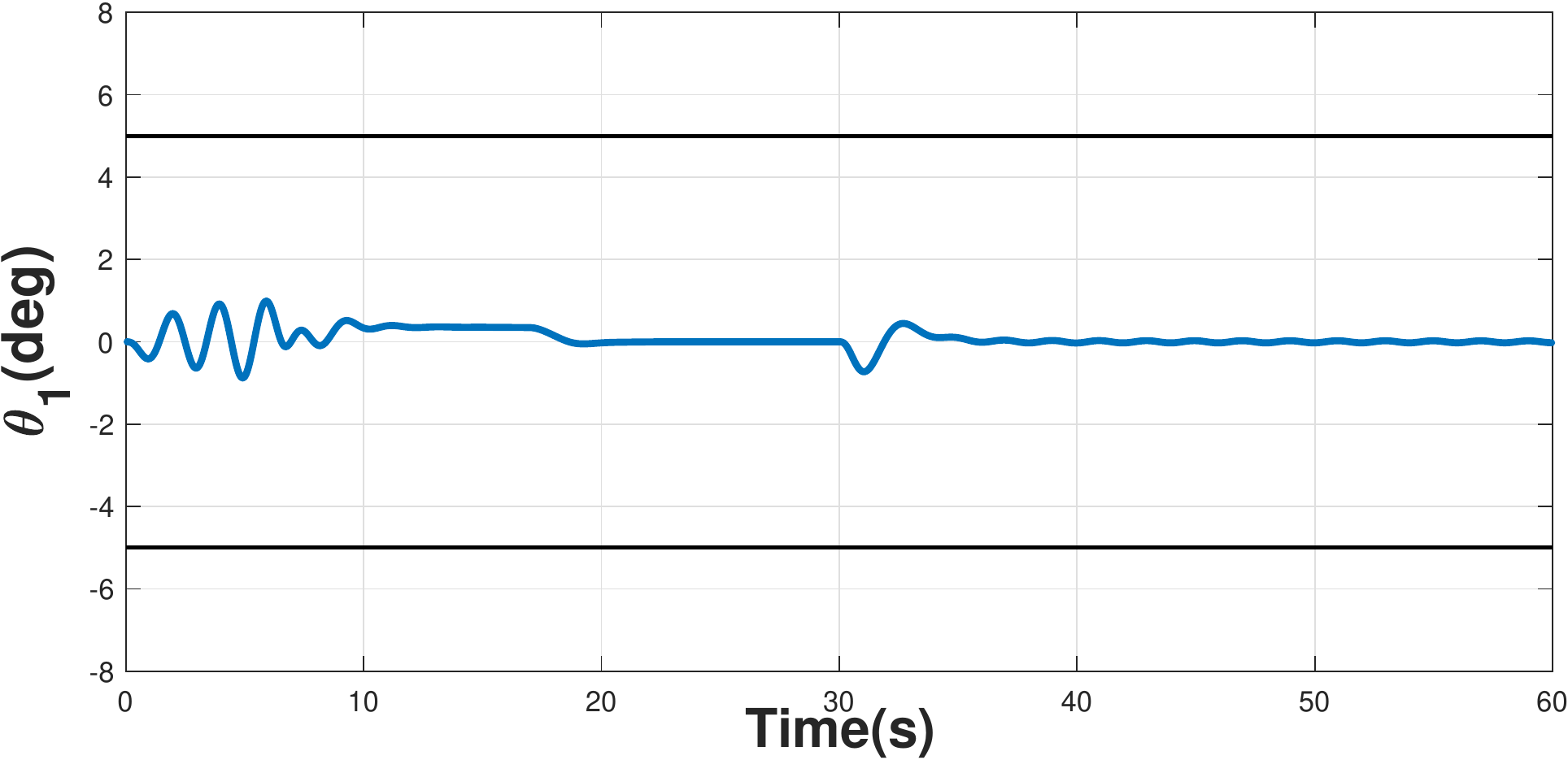}
\caption{\label{fig:t1} Payload angle  $\theta_1$. Blue line: Real value. Black line: Constraints}
\end{figure}

\begin{figure}[ht!]
\centering
\includegraphics[width=1\columnwidth]{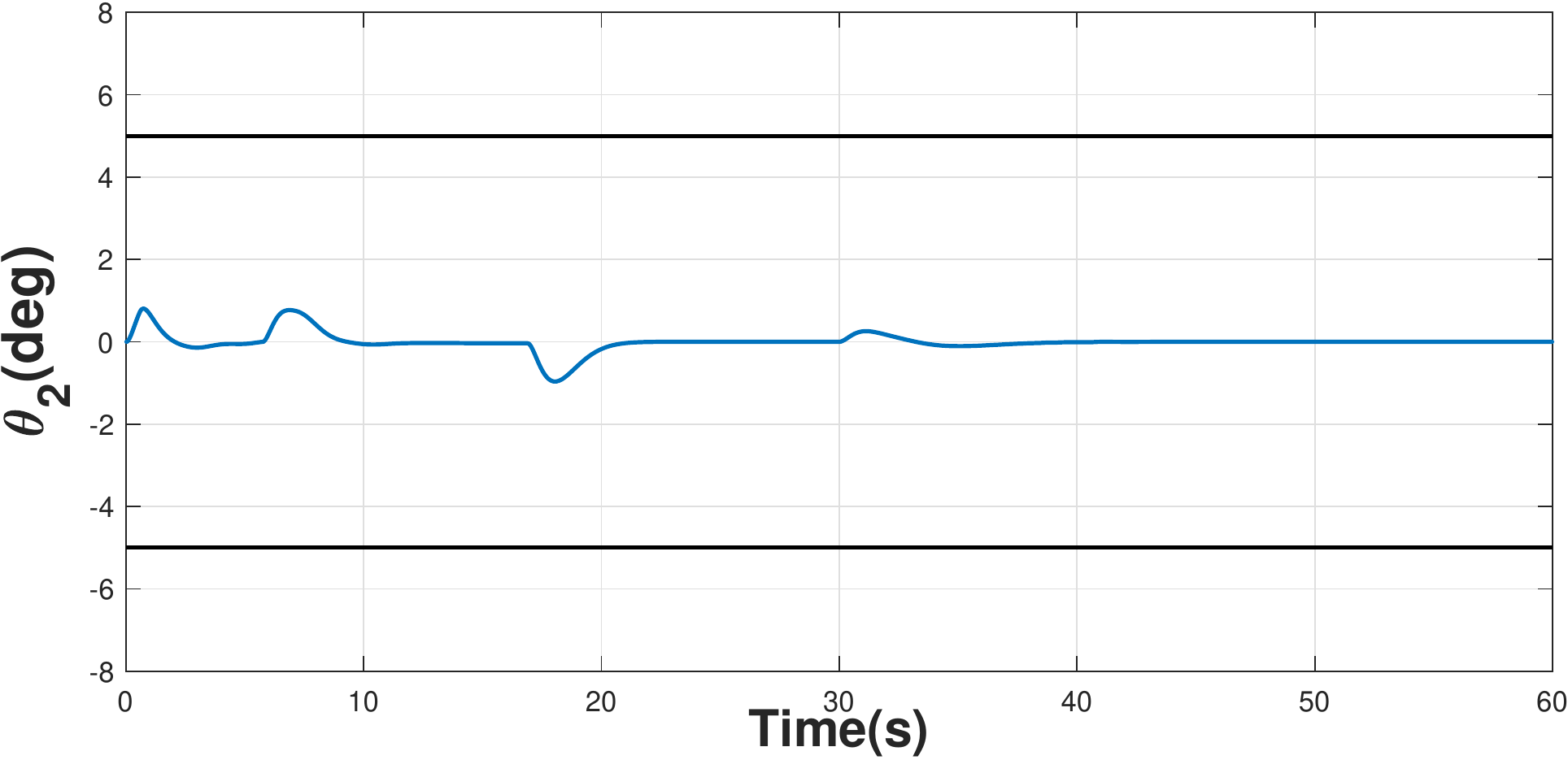}
\caption{\label{fig:t2} Payload angle $\theta_2$. Blue line: Real value. Black line: Constraints}
\end{figure}

\begin{figure}[ht!]
\centering
\includegraphics[width=1\columnwidth]{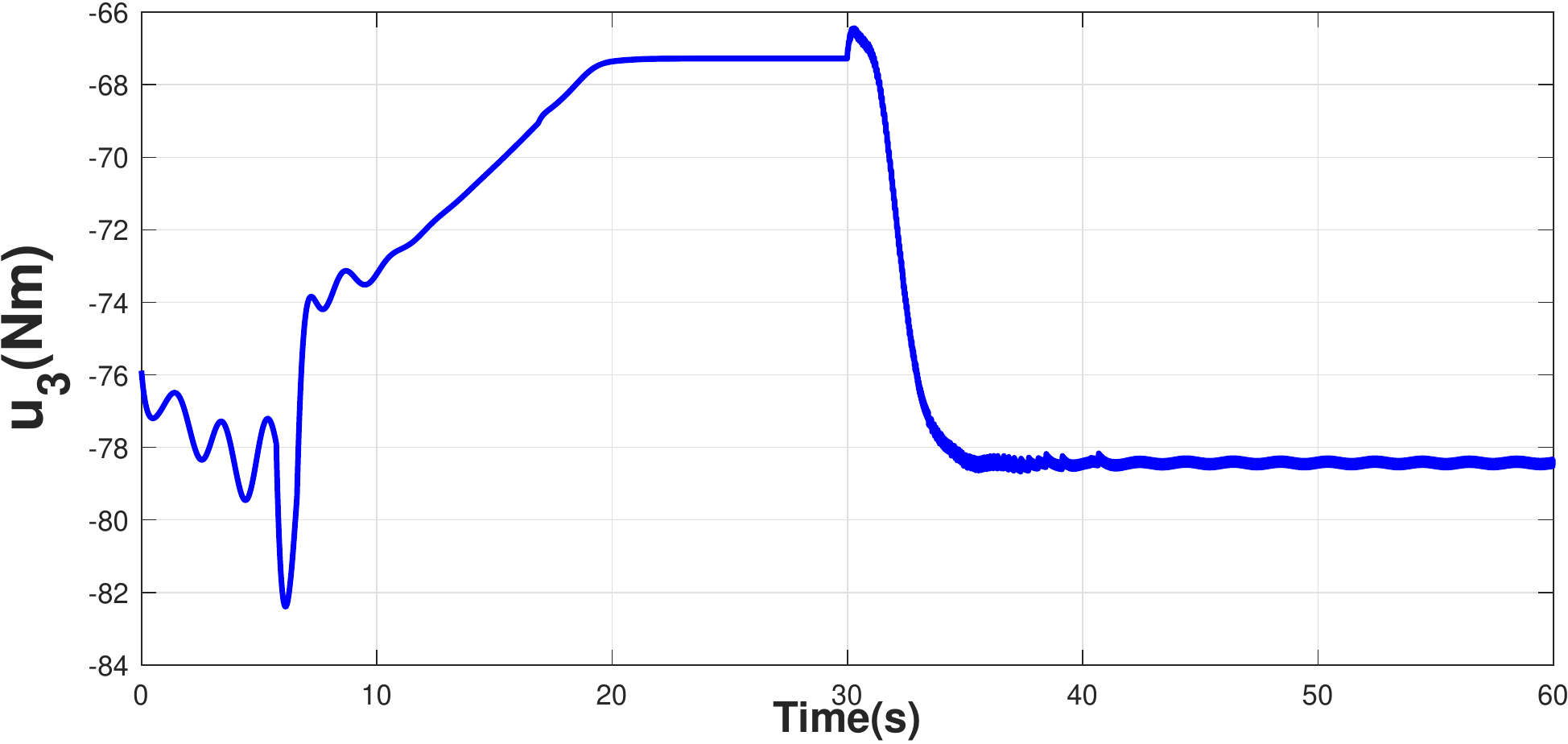}
\caption{\label{fig:u3} Input $u_3$}
\end{figure}

\begin{figure}[ht!]
\centering
\includegraphics[width=1\columnwidth]{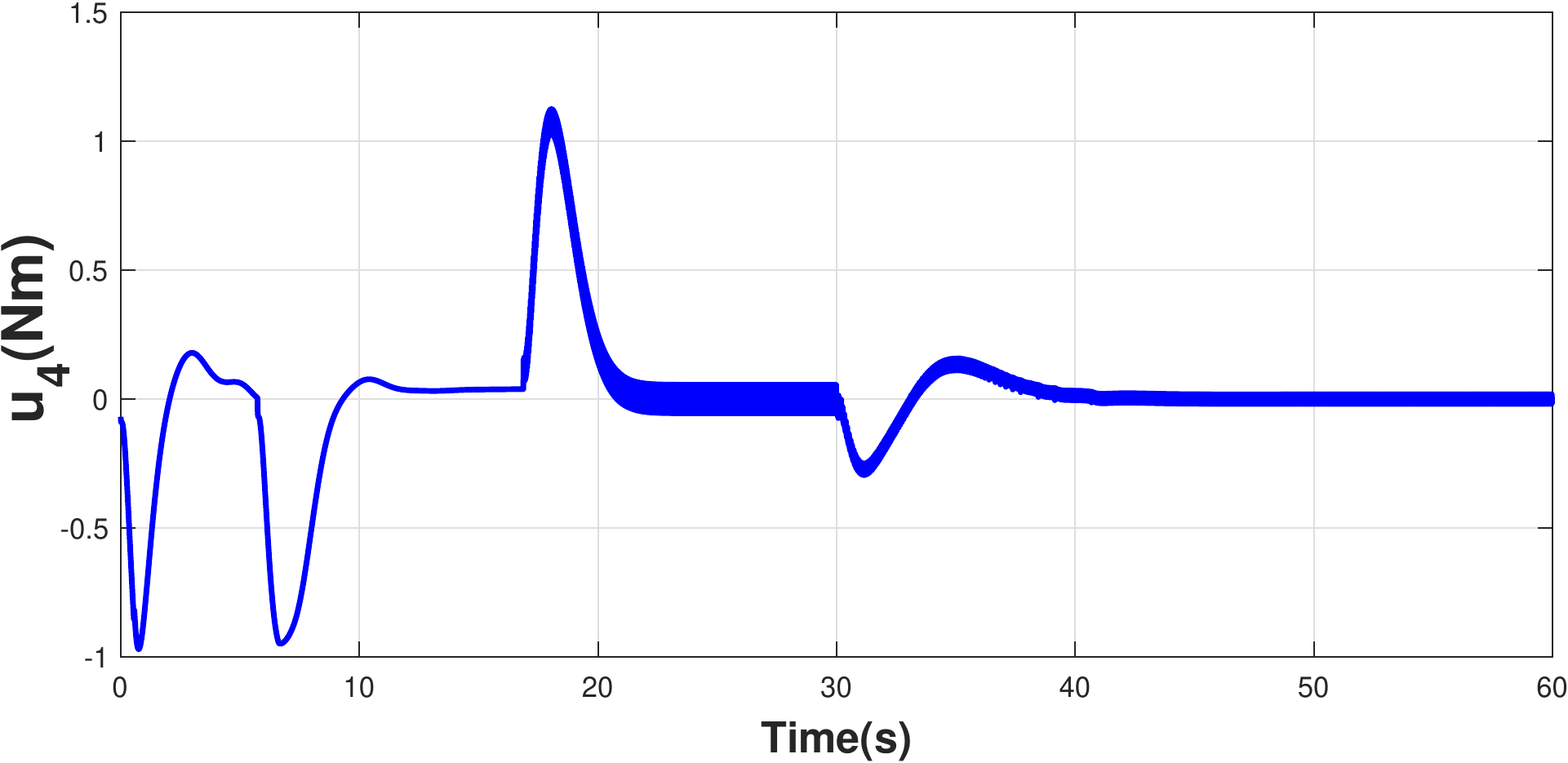}
\caption{\label{fig:u4} Input $u_4$}
\end{figure}

To validate the proposed scheme experimentally, we consider the following case scenario. Starting from the initial position $x_0=[0,0,{105\pi\over180},{\pi\over2},0,0,0,0]^T$. Similar to what happens in reality, we decided to first apply the first desired reference $[\theta_{3,r1},\theta_{4,r1}]=[{59\pi\over180},{-48\pi\over180}]$ to  move the crane over the mason. Later on, we apply the second desired reference $[\theta_{3,r2},\theta_{4,r2}]=[{88\pi\over180},{-58\pi\over180}]$ to move the payload in front of the mason.\newline
Figg.~\ref{fig:t3}-~\ref{fig:t4} show that the boom pitch and yaw angle follows the desired reference.\newline
As one can see, in Fig.~\ref{fig:ERG1} the system follows the desired reference and the control law is able to avoid collisions with the wall. In  Fig.~\ref{fig:ERG2} it is shown the same trajectory but in the space of end-effector.\newline
Figg.~\ref{fig:t1}-\ref{fig:t2} show that, during the desired trajectory, the payload swing angles do not violate the constraints (\ref{eq:constt11})-(\ref{eq:constt22}).\newline
It is worth noting that in this paper no constraints have been imposed on the actuation limits (see Figg.~\ref{fig:u3}-~\ref{fig:u4})). However, as one can seen from Figg.~\ref{fig:u3}-~\ref{fig:u4}, they do not represent a problem as the inputs profile and values are reasonable and well within the typical limits of the crane actuators.



\begin{figure}[ht!]
\centering
\includegraphics[width=1\columnwidth]{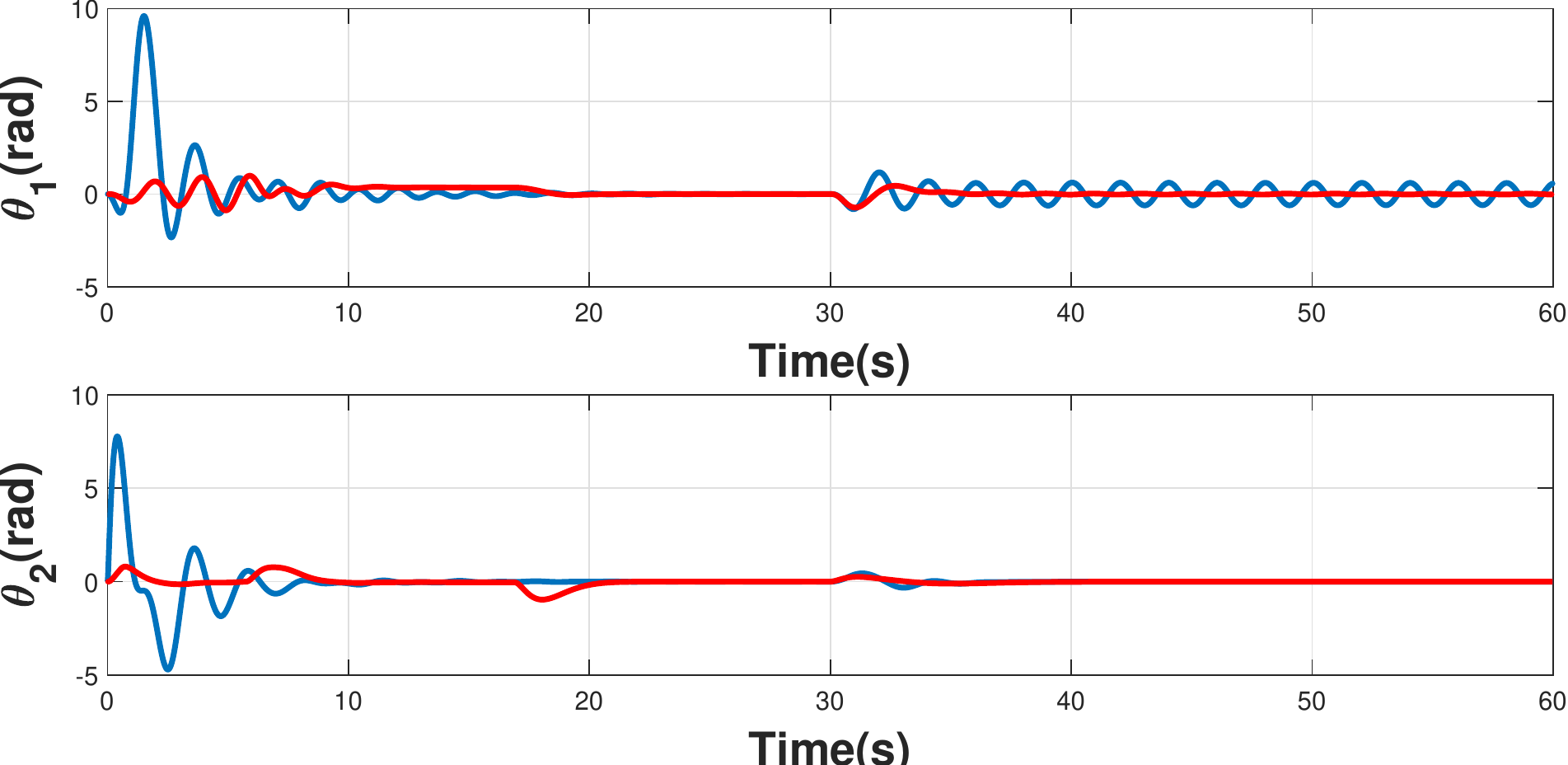}
\caption{\label{fig:conf_p} Payload swing angles. Red line: ERG. Blue line: Controller \cite{20}}
\end{figure}

\begin{figure}[ht!]
\centering
\includegraphics[width=1\columnwidth]{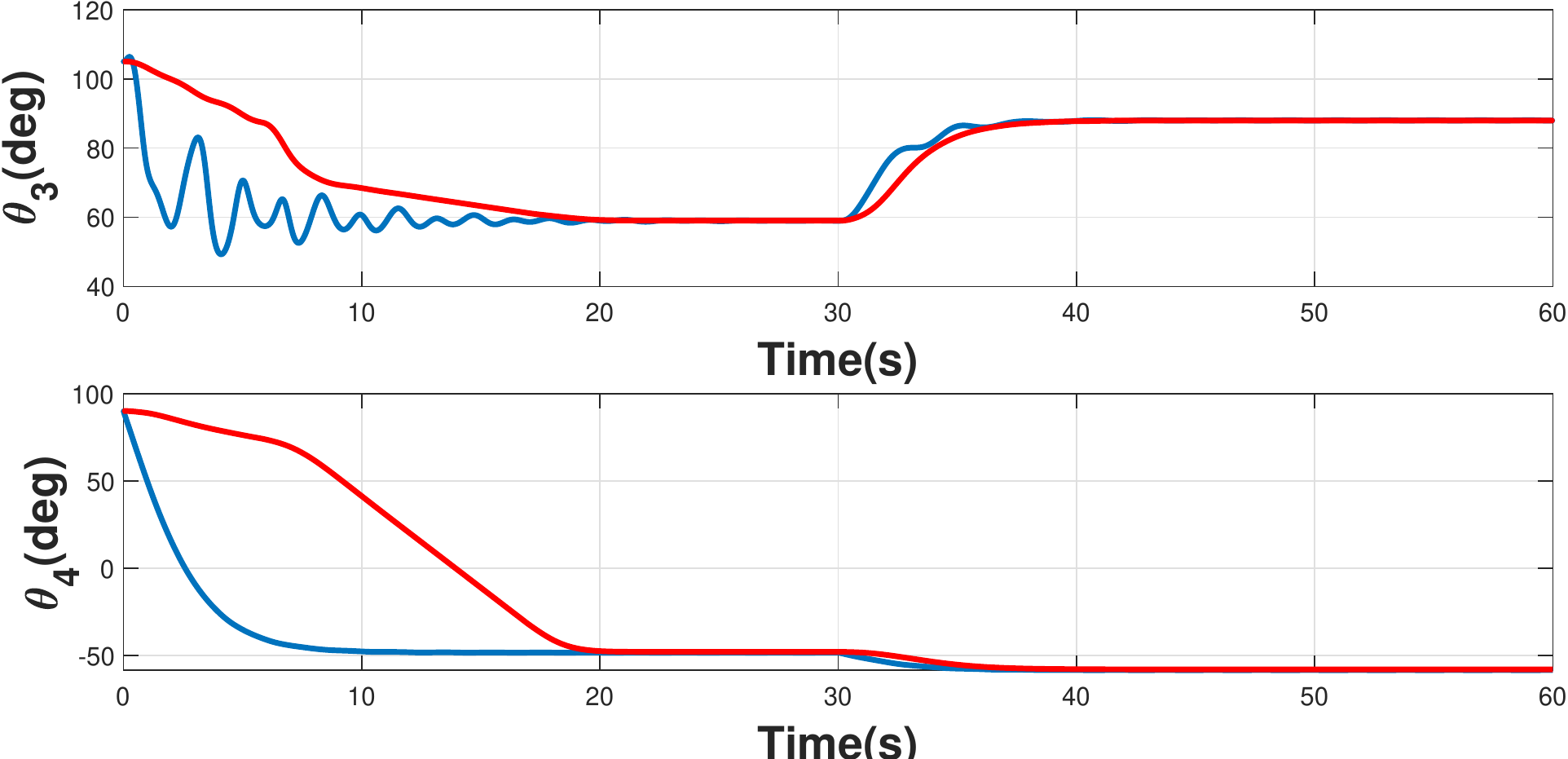}
\caption{\label{fig:conf_c} Boom angles. Red line: ERG. Blue line: Controller \cite{20}}
\end{figure}

Figg.~\ref{fig:conf_p}-~\ref{fig:conf_c} show the comparison between the proposed control strategy and the control reported in \cite{20}. As one can see, our controller is able to drive the crane to the desired position while the payload swings have a smaller amplitude.

\section{Conclusion}\label{sec:Conclusion}

This paper proposed a constrained control scheme based on the ERG framework for the control of boom cranes. The main contribution of this paper w.r.t. existing closed-loop control methods for boom cranes is that the proposed solution is able to guide the crane towards a desired reference, avoiding collisions with the wall and ensuring that the non-actuated variables (i.e.,$\theta_1$ and $\theta_2$) do not exceed a pre-defined maximum value. It is worth noting that no off-line trajectory has been calculated that it is the control law itself that decides how to move the reference to avoid the obstacle.


\bibliography{ISARC}

\end{document}